\documentstyle[epsf]{lamuphys}
\makeatletter
\let\chapter\hid@chapter
\makeatother
\begin{document}
\pagenumbering{arabic}
\title{Near-IR Properties of Quasar Host Galaxies}

\author{Kim K.\,McLeod\inst{}}

\institute{Smithsonian Astrophysical Observatory, 60 Garden St.,
Cambridge, MA 02140, USA}

\maketitle

\def\micron{\hbox{$\mu m$}}                     
\def\perpix{\hbox{$\rm pix^{-1}$}}              
\def\arcsec{\hbox{$^{\prime\prime}$}}           
\def\msa{\hbox{$\rm mag~arcsec^{-2}$}}          
\def\kms{\hbox{$\rm km~s^{-1}$}}                
\def\Hfifty{\hbox{$H_0=50\rm~km~s^{-1}~Mpc^{-1}$}}   
\def\Heighty{\hbox{$H_0=80\rm~km~s^{-1}~Mpc^{-1}$}}  
\def\q00{\hbox{$q_0=0$}}                          
\def\lstar{\hbox{$L^*$}}                          
\def\rquarter{\hbox{$r^{1/4}$}}                   
\def\lesssim{\mathrel{\hbox{\rlap{\hbox{\lower4pt\hbox{$\sim$}}}\hbox{$<$}}}}
\def\gtrsim{\mathrel{\hbox{\rlap{\hbox{\lower4pt\hbox{$\sim$}}}\hbox{$>$}}}}
\def\Msun{\hbox{$M_\odot$}} 
\def\Lsun{\hbox{$L_\odot$}} 
\def\etal{et al.}         

\begin{abstract}

We have obtained deep, near-IR images of nearly 100 host galaxies of nearby
quasars and Seyferts.  We find the near-IR light to be a good tracer of
luminous mass in these galaxies.  
For the most luminous quasars there is a correlation between the
maximum allowed B-band nuclear luminosity and the host galaxy mass,
a ``luminosity/host-mass limit''.
Comparing our images with images from HST, we find that the hosts of
these very luminous quasars are likely early type
galaxies, even for radio-quiet objects whose lower-luminosity
counterparts traditionally live in spirals.  We speculate that the
luminosity/host-mass limit represents a physical limit on the size of
black hole that can exist in a given galaxy spheroid mass.
We discuss the promises of NICMOS for detecting
the hosts of luminous quasars.

\end{abstract}
\section{Introduction}

Although most of the attention given to AGN is concentrated on the
``N,'' there are 
compelling reasons to understand the ``G.''  The central engine
and the host galaxy must influence each other, and 
the exact connections hold crucial clues for understanding the
quasar phenomenon.  Moreover, it is plausible that nuclear activity
has played a role in the evolution of a significant fraction of  all
galaxies; Seyferts account for $\gtrsim10\%$ of galaxies today (Maiolino \&
Rieke 1995; Ho 1996), and AGN were even more important in the past.
Therefore, to understand the evolution of galaxies, we must
understand the host galaxies of AGN.

By now it is well established that 
the redshift range $2\lesssim z \lesssim3$ represents a critical
period in the evolution of both ``normal'' galaxies and quasars.  
It is likely that galaxies at 
that epoch were starting to turn their gas into stellar disks. 
The mass in neutral hydrogen gas in damped Lyman$-\alpha$ absorbers at
that redshift is comparable to the mass in disk stars today and shows
strong evolution since that time 
(Lanzetta \etal~1995; Storrie-Lombardi \etal~1996).  Furthermore, a
photometric-redshift analysis of the Hubble 
Deep Field shows that the luminosity density from star-forming
galaxies peaks near that same redshift (Sawicki \etal~1996).  
The AGN luminosity function shows a similarly strong
evolution; if described in terms of luminosity evolution (but see Wisotzki
et al. in these proceedings), the
``characteristic'' luminosity of quasars increases as
$\sim(1+z)^{3.4}$ to $z\sim2$ and flattens between $2<z<3$ (Boyle
1993).  Thus {\it quasars
were an energetically important component of  galaxies at a critical
period in their evolution and could have influenced their growth.}
Understanding the relationships between AGN and their hosts thus holds clues to the
processes of galaxy formation and evolution.

\begin{figure}
\epsfsize=0.45\vsize
\vbox{
\epsfbox{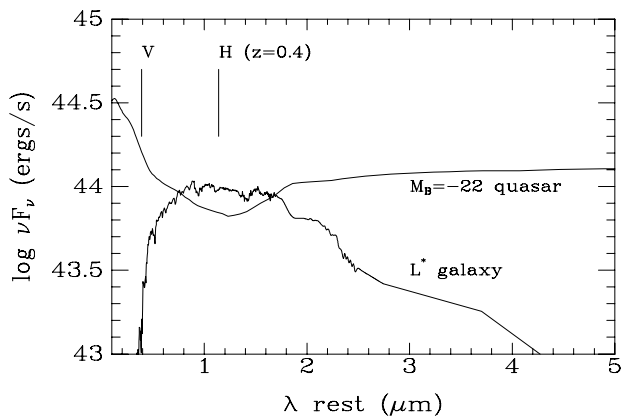}
}
\caption{Energy distributions of a typical quasar (from J. McDowell)
and galaxy (from M. Rieke) for $\rm H_0=80$.  The H-band gives good contrast of
starlight to nuclear light and is a good tracer of luminous
mass in the host galaxy.}\label{fig-seds}
\end{figure}

HST and near-IR imaging together have become
especially powerful tools for studying quasar hosts.
The spatial resolution of HST has allowed us, for the first time, to
determine morphological types for hosts of luminous quasars (McLeod
and Rieke 1995b; Bahcall et al. and Disney et al. in these proceedings).
Near-IR imaging has done a good job of showing the
host galaxy starlight with less contamination from the
nucleus than at visible wavelengths (McLeod \& Rieke
1994ab,1995ab; Dunlop \etal~1993; Kotilainen \& Ward 1994).
As shown in Fig. \ref{fig-seds}, the host galaxy starlight peak
coincides with the near-IR minimum in the nuclear energy distribution.
Furthermore, the near-IR light highlights the old, red, mass-tracing
stars in the population while suffering less from extinction and
emission-line contamination than the visible images.

\section{The IR Images}

We have exploited this wavelength range by 
obtaining deep near-IR images for nearly 100 AGN using a 256x256
NICMOS array camera on the Steward Observatory 2.3m telescope.
We chose two samples of AGN that allowed us to investigate host galaxy
properties over 10 
B magnitudes in nuclear luminosity.  For low-luminosity
AGN we used the CfA Seyfert sample, which is
selected on the basis of the nuclear spectrum and which has roughly
equal numbers of Sy1's and Sy2's.  For high-luminosity AGN, we chose
the lowest redshift ($z<0.3$) quasars from the PG sample, to ensure a
sample selected on the basis of nuclear 
properties and close enough so that the host galaxies would be
resolved.  The quasars were imaged in the H-band (1.65\micron) and
we were able to measure the isophotes down to a $1\sigma$ level of 
$\rm H\approx23\msa$, which corresponds approximately to $\rm B\approx
26.7\msa$ for typical galaxy colors.

The results have been presented in McLeod and Rieke (1994ab,1995ab).
We describe in these papers the 
relationships between host galaxy luminosities (and masses) and
nuclear properties; the existence of substantial obscuration coplanar
with the disks of host spirals of Seyferts; and the search for signs
of disturbances that could aid the flow of fuel towards the centers of
the galaxies.  The reader is encouraged to consult
these papers for details; here we will concentrate on 
one of the most intriguing results.

\section{The Luminosity/Host-Mass Limit: Observations}

\begin{figure}
\epsfsize=0.6\vsize
\vbox{
\epsfbox{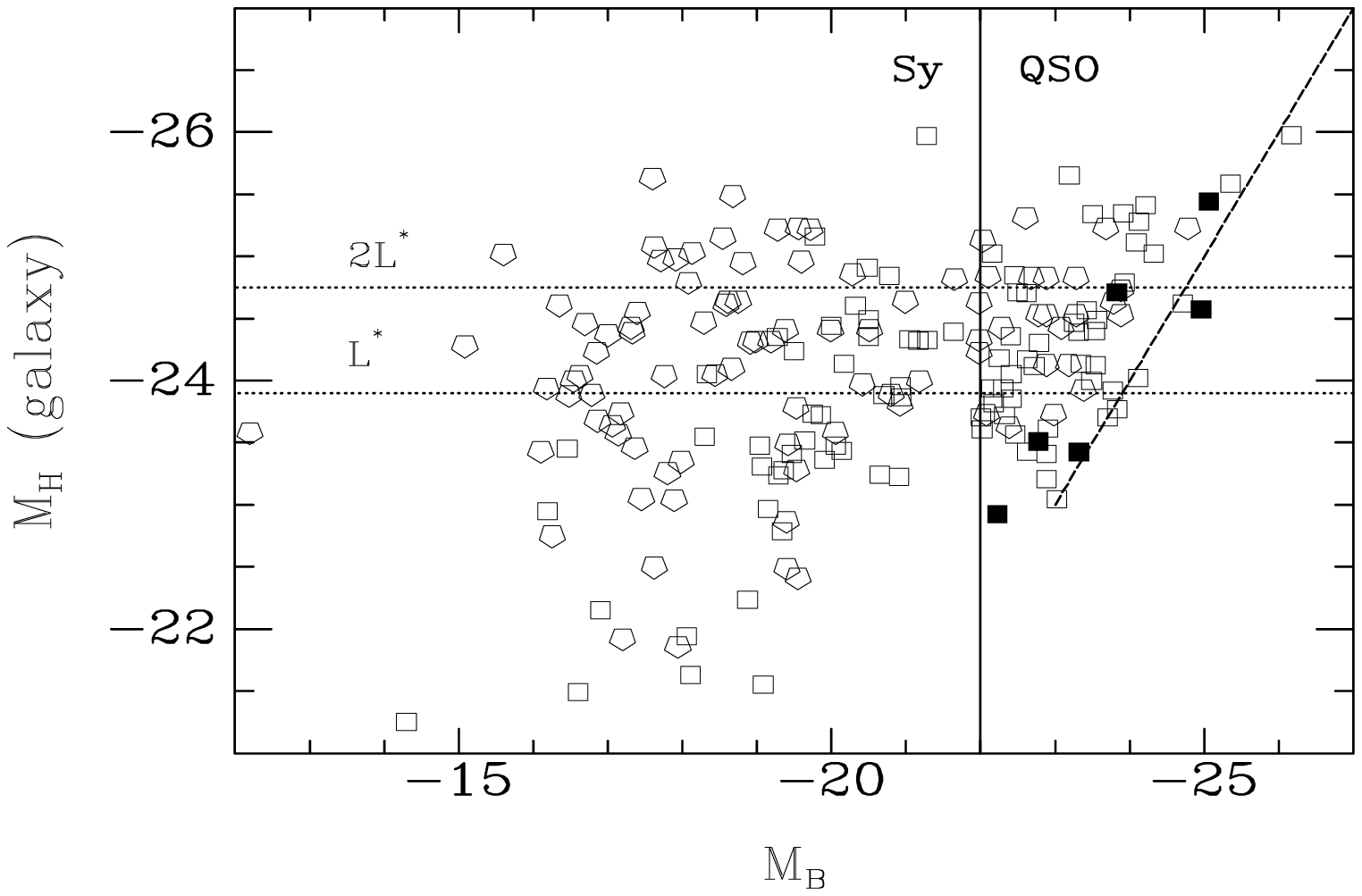}
}

\caption{Host v. nuclear luminosity for Seyferts and quasars
with $z<0.3$ for $\rm H_0=80$}\label{fig-absmags}
(McLeod \& Rieke 1995a and references therein).
For quasars brighter than $\rm M_B\lesssim -23$, there is a minimum host
galaxy luminosity that increases with nuclear power.  The dotted lines
show the positions of \lstar~and 2\lstar~galaxies.  Filled squares are
objects for which the host galaxy luminosities are upper limits or
weak detections--we will obtain NICMOS images for these and other
high-luminosity objects.  The diagonal line shows the 
luminosity/host-mass limit.

\end{figure}

We find the near-IR light to be a good tracer of
luminous mass in these galaxies.  In Fig. \ref{fig-absmags} we plot
host near-IR luminosity against nuclear luminosity, for the objects in
our samples and from other near-IR studies in the literature.
As shown in this figure, Seyferts are found in 
galaxies with a range of H-band luminosity, and hence luminous mass,
from $\sim 0.1 \rm ~to~\sim5\lstar$.  Their morphological types range
from S0 to Sc.  
The lowest-luminosity quasars live in similar kinds of galaxies 
spanning the same range of H-band luminosity centered around \lstar.

We have found, however, that for the
highest-luminosity quasars, there is a {\it minimum host
H-band luminosity} that increases with nuclear power, shown as a
diagonal line in Fig. \ref{fig-absmags}.  This is reminiscent of a 
similar trend previously noted in the visible (Yee 1992).  The
relationship has the functional form  $\rm 
M_H(galaxy)\approx M_B(nucleus)$ with an onset approximately one
magnitude brighter than the traditional (arbitrary)
quasar/Seyfert boundary.  Because  H-band light probes
the galaxy {\it mass} by highlighting the old,
red, mass-tracing stars of the stellar population, this 
``luminosity/host-mass limit'' reflects a relation between
fundamental physical parameters that govern the process of hosting a quasar:
galaxies with more mass can sustain more 
activity.

WFPC2 images have been published for $\sim30$ luminous, nearby quasars
including many of our highest-luminosity objects 
(e.g. Hutchings \etal~1994; Disney
\etal~1995; Bahcall \etal~1996ab).  Of these, only a few ($\sim3$) appear to
be spiral hosts like those of Seyferts.  The rest 
are plausibly smooth, early-type galaxies or 
ellipticals in the making (mergers), despite being radio quiet objects
that are traditionally assumed to be spirals (McLeod
\& Rieke 1995b; see also Taylor \etal~1996).

\section{The Luminosity/Host-Mass Limit: Wild Speculation}

Intriguingly, the transition from spiral hosts to spheroid-dominated
hosts appears to occur approximately at the nuclear luminosity where the
luminosity/host-mass limit becomes apparent. A simple explanation 
is that the objects along the diagonal line in Fig. \ref{fig-absmags}
are early-type 
galaxies that have a maximum allowed black hole mass for their galaxy
mass and that the black hole is accreting at the Eddington rate.  
In this case and for typical galaxy 
mass-to-light ratios, the diagonal line then 
represents a line of {\it constant fraction of black hole mass to
stellar spheroid mass}, with value $f_{BH}\equiv M_{BH} / M_{stars}
\approx 0.0015$. 

This suggestion is especially exciting because 
central compact objects (presumably dead quasars) discovered in nearby
spheroids show a  similar relation with $f_{BH}\approx 0.002$ (Kormendy
\& Richstone 1995), and an
HST study of bulges and ellipticals indicates that compact objects 
following approximately this same relation are likely required to 
produce the cores seen in the starlight (Faber \etal~1996).
Thus, the luminosity/host-mass limit possibly results from 
general physical processes that govern the formation and evolution of
the spheroid components of galaxies. 

\section{The Luminosity/Host-Mass Limit: Wilder Speculation}

If the luminosities of most powerful quasars really do trace the
potentials of the most massive spheroids, 
then we might be able to use the evolution of the quasar luminosity
function to trace the history of spheroid formation in the universe.
The observed evolution of the quasar luminosity function can be nicely
reconciled with the model of galaxy evolution recently summarized by 
Fukugita \etal~(1996), in which spheroid components of
very massive galaxies formed at $z>3$, followed by less massive objects
at later epochs (see also Sawicki \etal~1996).   
We speculate that the $z\sim5$ quasars 
formed as parts of the most massive spheroids, the 
peak in the quasar population at $z\sim2$
occurred when massive disks were being added, and less powerful AGN formed
later in less massive galaxies.  Thus, {\it central black holes
themselves form as the spheroids are being assembled, and the changing
AGN population reflects the changing population of galaxies.}
The luminosity functions would then require nearly every large
galaxy to go through an AGN phase that lasts a modest portion of the
galaxy's lifetime (e.g. Weedman 1986). 

\section{Future Work: All Hail NICMOS!}

Before we can exploit the luminosity/host-mass limit we must
answer several questions.  We are already at work on the theoretical
basis for such a limit, and we are 
beginning a groundbased imaging and spectroscopy program to 
determine whether the limit applies to AGN in smaller 
spheroids (Seyferts in bulges). 

\begin{figure}
\epsfsize=0.45\vsize
\vbox{
\epsfbox{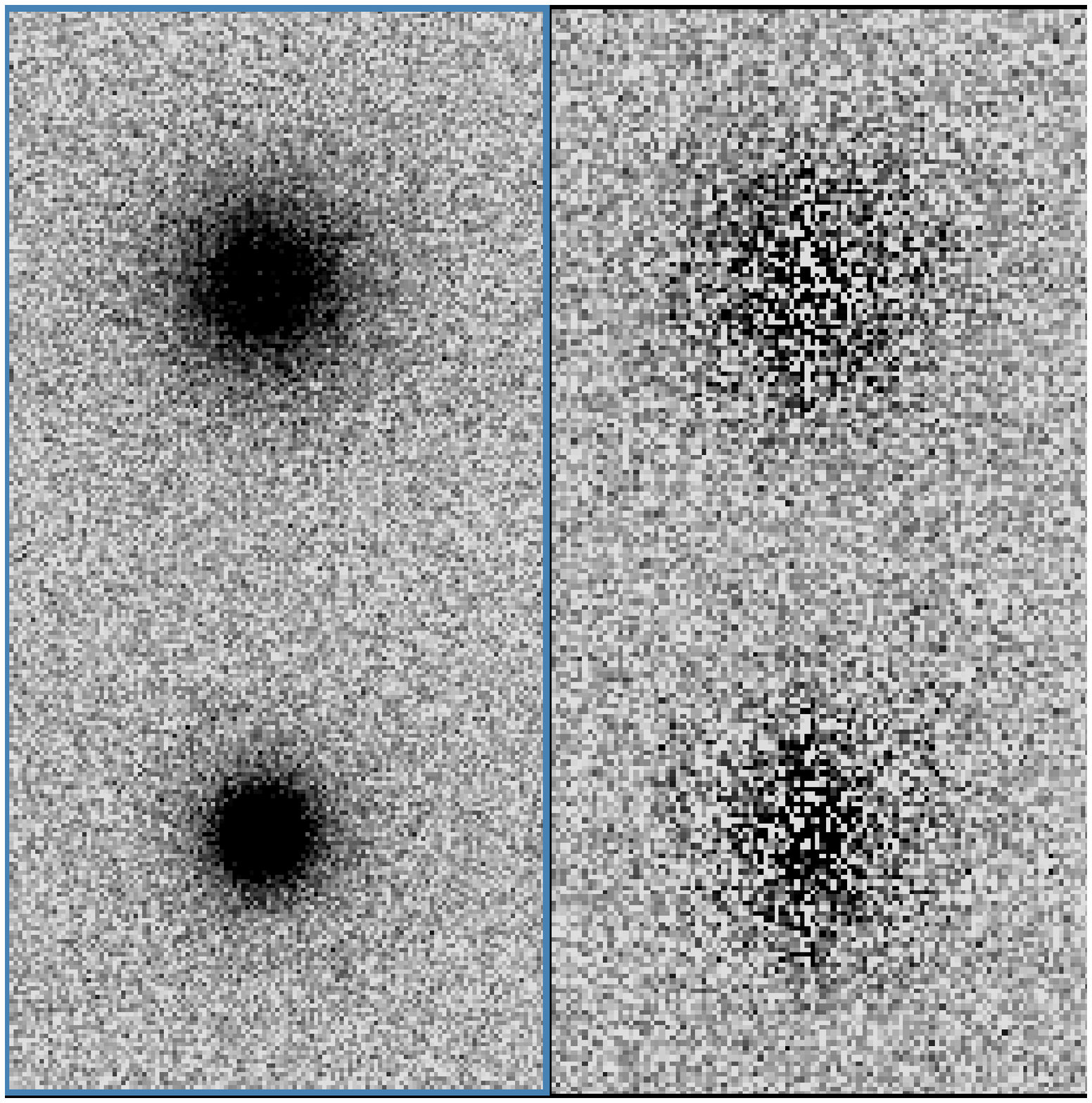}
}

\caption{Simulated 2800s images of $z=0.4$ quasars with NIC2
(left) and WFC2 (right).}\label{fig-sims}
The hosts at top are \lstar~spirals, the ones at bottom are
\lstar~ellipticals.  In both cases, the nucleus has been removed and
the linear greyscale stretch runs from -1 to 10 times the 1$\sigma$
noise.  The frame is 19.2'' on a side (the size of one NIC2 frame).

\end{figure}

Importantly, we need to know whether the limit is strictly correct
especially at high nuclear luminosities where our statistics are poor.
We have been granted Cycle 7 HST time to address this question with
NICMOS.  NICMOS will combine the advantages of near-IR imaging with
the superior spatial resolution of HST.
We will look to find exceptions to the luminosity/host-mass limit by imaging 
objects whose hosts have been thus far elusive from the ground.  We
will also extend our sample to slightly higher redshifts to allow a
look at higher luminosity objects.  As shown in Fig. \ref{fig-sims},
NICMOS will be very good at detecting the extended emission from
smooth hosts that are difficult to see with WFPC2.

Eventually, we need to probe nucleus--host relationships at high
redshifts where galaxies are being assembled.  
The NICMOS GTO program (Weymann \etal) will
provide near-IR images of radio-quiet quasar hosts at $z\approx1$.
We will probe further by obtaining near-IR images using adaptive
optics on the soon-to-be-upgraded MMT.  Preliminary adaptive optics
results (John Hutchings, this conference) suggest that this method 
holds great promise for host studies at high redshift.  By measuring 
morphologies and near-IR luminosities of the distant quasar hosts from
the images and then applying models of stellar and dynamical
evolution, we will predict what those galaxies look like today.

\medskip
Acknowledgements:  Thanks to the conference organizers for conference
organizing, thanks to Avi Loeb for very fruitful discussions, 
thanks to Dr. Yamada for pointing out the Kormendy ratio, and thanks
to B. Wilkes for financial support.

%

%
%

\end{document}